# Quantum supercurrent transistors in carbon nanotubes

Pablo Jarillo-Herrero, Jorden A. van Dam, Leo P. Kouwenhoven

*Kavli Institute of Nanoscience, Delft University of Technology, PO Box 5046, 2600 GA, Delft, The Netherlands*

**Electronic transport through nanostructures is greatly affected by the presence of superconducting leads[1-3]. If the interface between the nanostructure and the superconductors is sufficiently transparent, a dissipationless current (supercurrent) can flow through the device due to the Josephson effect[4,5]. A Josephson coupling, as measured via the zero-resistance supercurrent, has been obtained via tunnel barriers, superconducting constrictions, normal metals, and semiconductors. The coupling mechanisms vary from tunneling to Andreev reflection[5-8]. The latter process has always occurred via a normal-type system with a continuous density of states. Here we investigate a supercurrent flowing via a discrete density of states, i.e., the quantized single particle energy states of a quantum dot[9], or *artificial atom*, placed in between superconducting electrodes. For this purpose, we exploit the quantum properties of finite-sized carbon nanotubes[10] (CNTs). By means of a gate electrode, successive discrete energy states are tuned ON and OFF resonance with the Fermi energy in the superconducting leads, resulting in a periodic modulation of the critical current and a non-trivial correlation between the conductance in the normal state and the supercurrent. We find, in good agreement with existing theory[11], that the product of the critical current and the normal state resistance becomes an oscillating function, in contrast to being constant as in previously explored regimes.**

In artificial atoms current can flow via discrete states according to the general process of resonant tunneling, i.e., *resonant* when the Fermi energy in the leads is aligned with discrete energy states[9]. The maximum conductance, $G$, through a single spin-degenerate energy level depends on the coupling to the leads. For the case of phase-coherent tunneling, $G$ can reach $2e^2/h$, when charging effects are unimportant. If charging effects are significant, still $G=2e^2/h$ can be achieved (even off-resonance) by means of the Kondo effect[12], which establishes spin coherence between the quantum dot (QD) and the leads. An entirely new situation arises in the case of superconducting leads, i.e., when two superconductors are coupled via a discrete single particle state. As we show below, the conductance can reach infinity, that is, a supercurrent can flow through the QD. This means that $G$ exceeds by far the perfect conductance level of $2e^2/h$ occurring when the transmission probability reaches one. This zero resistance state is peculiar since just a single discrete state, that can be occupied only with two spin degenerate electrons simultaneously, is available for coupling the collective macroscopic states in the leads. In contrast to previously accessible regimes, we can study Josephson coupling for ON and OFF resonant tunneling, which enables a transistor-like control of the supercurrent through the quantum dot.

The carbon nanotube devices are fabricated by means of standard nanofabrication techniques and geometries (e-beam lithography to define customized electrodes on CNTs grown by chemical vapour deposition on top of oxidized silicon substrates[13]) with two extra important ingredients: the choice of superconducting material and a multiple-stage filtering system to suppress electronic noise over a wide frequency range (see Fig. 1a and supplementary information for details).

The quantum behaviour of electrons in carbon nanotubes in good contact with metallic electrodes emerges clearly in a measurement of the differential resistance, $dV/dI$, versus measured source-drain voltage, $V$ and gate voltage, $V_G$, as shown in Fig. 1b for one of our devices in the normal state. The differential resistance exhibits a pattern of high and low conductance regions, typical of nanotube devices well coupled to the leads[14, 15], with a characteristic voltage scale, $V$~3.5 mV. This energy corresponds to the energy level separation between the discrete electronic states due to the finite length of the CNT, $\Delta E = h v_F / 2L$, where $h$ is Planck's constant, $v_F = 8.1 \cdot 10^5$m/s is the Fermi velocity in the CNT, and $L$ its length. The value obtained from this measurement, $L$~480nm, is in good agreement with the length of the nanotube segment in between the metallic electrodes, 470nm. When the sample is cooled down below the superconducting critical temperature of the electrodes (~1.3K), the electronic transport through the nanotube is strongly affected due to the superconducting proximity effect[1, 16-19], which can be viewed as the leakage of Cooper pairs from a superconductor into a normal metal-type material. This proximity effect is evident from the observation of multiple Andreev reflections (MAR)[20] and the flow of a supercurrent through the device (Figs. 1c,d). We note that we have observed similar supercurrents in 4 out of 7 measured metallic CNT devices with room temperature resistances below 35kΩ (see supplementary information for additional data and magnetic field dependence). The most interesting feature of this supercurrent is that its maximum value (critical current, $I_C$) can be strongly modulated by means of a gate electrode[21], as shown in Fig. 1d. Since the CNTs are metallic, this means that the supercurrent transistor action must have a different mechanism than in conventional semiconductor structures. It is also remarkable that the gate voltage necessary to change from maximum to minimum $I_C$ is of only ~50 mV, much smaller than the typical gate voltages necessary to significantly vary

4the charge density of semiconducting carbon nanotubes[22] or nanowires with similar geometries[23].

In order to establish the origin of the modulation of $I_C$, it is important to characterize the sample over a larger gate voltage range. A measurement of $dV/dI$ $(I,V_G)$ (Fig. 2b) shows a non-monotonic, quasi-periodic set of low differential resistance regions, where $I_C$ is largest, in between regions of high $dV/dI$, where $I_C$ is strongly suppressed (Fig. 2a). This pattern follows closely the low-bias pattern of Fig. 1b, with the same gate voltage spacing in between resonances, but now the vertical axis is current, instead of voltage. The correspondence between the two patterns indicates that the modulation of $I_C$ is due to the tuning ON and OFF resonance with gate voltage of the energy levels in the CNT with respect to the Fermi energy in the leads (as shown schematically in Fig. 2e). Such Josephson transistor mechanism, purely due to the discrete nature of the energy levels in a nanostructure (in this case finite-sized CNTs), has not been previously observed.

Before turning to a more quantitative description, we note that the modulation of $I_C$ is followed by a series of $dV/dI$ peaks and dips moving up and down in the current axis. These are better seen in the high-resolution measurement shown in Fig. 2c and reflect the multiple Andreev reflection processes (see also Fig. 1c) taking place at the CNT-metal interfaces. MAR processes occur at voltages $V = 2\Delta_g/en$ ($\Delta_g$ is the superconducting energy gap, $e$ is the electron charge, $n$ an integer number). The $dV/dI$ curves in fact occur at constant voltage in this current-biased sample (see supplementary information). Two individual $dV/dI$ traces are shown in Fig. 2d for the ON and OFF resonance situations. In both cases $dV/dI$ exhibits oscillations due to MAR, but the overall behaviour of $dV/dI$ is very different. In the ON resonance case, $dV/dI$ *decreases* with *decreasing* $|I|$ (on average) until it "switches" to zero when ~$I_C$ is reached. For the OFF case, $dV/dI$ *increases* with *decreasing* |*I*| (except at the MAR points), until |*I*| reaches a very strongly suppressed value of $I_C$ (barely visible in Fig. 2d, see Fig. 3). From the normal state resistance values in the ON and OFF resonant cases, we can conclude that the *dV/dI* changes between these qualitative behaviours at values of $dV/dI \sim h/(2e^2) \sim 13$ kΩ, i.e., once the resistance per channel of the CNT becomes of the order of the quantum of resistance (see also supplementary information).

The correlation between the critical current and the normal state resistance, $R_N$, is well studied in S-"normal metal"-S (SNS) structures. As a matter of fact, for short junctions in diffusive systems and ideal NS interfaces, $I_C R_N \sim \Delta_g/e$, i.e. constant[5]. The situation differs when one considers a single discrete energy level. In this case, the conductance is given by $G_N = (4e^2/h)T_{BW}$, where $T_{BW} = \Gamma_1\Gamma_2/((\varepsilon_R/h)^2 + 0.25\Gamma^2)$ is the Breit-Wigner transmission probability, $\Gamma_{1,2}$ are the tunnel rates through the left/right barriers ($\Gamma=\Gamma_1+\Gamma_2$), and $\varepsilon_R$ is the energy of the resonant level relative to the Fermi energy in the leads. (Note that we have added a factor of 4 in $G_N$ to account for the spin and orbital degeneracy of the CNT electronic states[10, 15, 24, 25].) Beenakker and van Houten[11] have studied the lineshape for the critical current in such a system. (The general case, including finite length effects has been studied by Galaktionov and Zaikin[26].) For the case of a wide resonance, $h\Gamma \gg \Delta_g$, they obtained $I_C = I_0 [1-(1-T_{BW})^{1/2}]$, with $I_0 = 2e\Delta_g/\hbar$. Experimentally, we can vary the position of the resonant level by means of a gate voltage, $\varepsilon_R \propto V_G$, as shown for the normal state conductance in Fig. 3b. From the maximum value of $G_N \sim 3.8\ e^2/h$, we deduce a barrier asymmetry $\Gamma_1/\Gamma_2 \sim 0.64$. We use this to fit $I_C (V_G)$ and $G_N(V_G)$ (see Fig. 3a,b; red curves). Although the functional form is in good agreement with theory, the values for $\Gamma$, $\Gamma_I = 0.85$ meV/$h$ and $\Gamma_G = 1.36$ meV/$h$, obtained from the $I_C (V_G)$ and $G_N(V_G)$ fits, respectively, differ substantially. Also the value of $I_0$ determined, 4.15nA, is much smaller than the theoretical value ($I_0=2e\Delta_g/\hbar \sim 60$nA, resulting in a maximum theoretical supercurrent of 47nA when the asymmetry in the barriers is taken into account[11]). Although finite length effects can yield a partial reduction of $I_C$ (ref. 26), they cannot explain the drastic suppression (a factor ~15) that we measure. Such strong suppression is reminiscent of the behaviour of small, underdamped, current-biased Josephson junctions[27], where the electromagnetic environment leads to a measured critical current, $I_{CM}$, much lower than the true critical current $I_C$ (see supplementary information for an estimate of the quality factor and suggestions for increasing $I_{CM}$). The dynamics of such a Josephson junction can be visualized as that of a particle moving in the so-called "tilted washboard" potential[5], where the driving current corresponds to the tilt in the potential. For the case of low dissipation (underdamped junctions), a small fluctuation can cause the particle to slide down the potential and go into a "runaway" state. This occurs at a value of $I$ much smaller than the true $I_C$, and it has been shown[27] that the measured critical current scales as $I_{CM} \propto (I_C)^{3/2}$. In order to test the applicability of this model to CNT Josephson junctions, we have fitted $I_{CM} = I_{0M} [1-(1-T_{BW}(V_G, \Gamma))^{1/2}]^{3/2}$, as shown by the blue curve in Fig. 3a. We obtain a similarly low value of $I_{0M} = 4.57$nA and this time, the value of $h\Gamma_I$ obtained, 1.22 meV, is in good agreement with $h\Gamma_G = 1.36$meV, resulting also in an improved fit to the data.

The importance of the coupling to the environment manifests itself more explicitly when examining the correlation between the critical current and the normal state conductance. We note that in the case of an ideal diffusive SNS junction the correlation would yield a simple straight line. The experimental data severely deviate from such curve (Fig. 3c). First we consider the expected theoretical decay for the case of a discrete state (red curve) $I_C = I_0 [1-(1-¼G_N)^{1/2}]$ (no fitting parameters), with the value of $I_0$ obtained from Fig. 3a, and $G_N$ measured in units of $e^2/h$. The comparison with the predicted theoretical line shows that the measured $I_C$ is significantly lower than expected. However, a remarkably better agreement is found when the electromagnetic environment is included, as shown by the blue curve, $I_{CM} = I_0 [1-(1-¼G_N)^{1/2}]^{3/2}$, indicating the generality of the $(I_C)^{3/2}$ dependence of $I_{CM}$ for very different type of Josephson junctions[27].

The predicted lineshape of $I_C$ (even in the presence of low dissipation) implies that the $I_C R_N$ product is not constant, but instead has a maximum on resonance. We plot in Fig. 3d the $I_C R_N$ product, which indeed exhibits a peak structure. The red and blue lines, which contain no extra fitting parameters, result from dividing the theoretical curves in Fig. 3a by the red curve in Fig. 3b, and further substantiate the results from earlier figures.

We emphasize that the above-mentioned analysis confirms the correct order of the relevant energy scales necessary for the observation of the resonant tunneling supercurrent transistor action[11]: $\Delta E$ (~ 3.5 meV) > $h\Gamma$ (~ 1.3 meV) >> $\Delta_g$ (~ 125 μeV) > $U$. The last inequality is justified since signatures of Coulomb blockade effects, such as a four-fold splitting of the conductance peaks, are absent in our data, concluding that the charging energy, $U$, is negligible (see supplementary information).

We end by noting that, although both superconductivity and the Kondo effect are collective many-body phenomena, their effect on resonant tunneling is very different[18]. The Kondo enhancement occurs OFF-resonance, while the superconducting zero-resistance state, as we have shown, is most pronounced ON-resonance. In fact, we expect that the study of CNT devices with intermediate transmission, and thus, larger Coulomb interactions, will enable the observation of Kondo-enhanced supercurrents in the OFF resonant case[28, 29].

**Supplementary Information** accompanies the paper on **www.nature.com/nature**.

**Author Information:** Reprints and permissions information is available at

npg.nature.com/reprintsandpermissions. The authors declare no competing financial interests.

Correspondence and requests for materials should be addressed to P.J. (e-mail: Pablo@qt.tn.tudelft.nl)

We thank Yu. V. Nazarov, C. W. J. Beenakker, W. Belzig, S. De Franceschi and Y-J. Doh for discussions and C. Dekker for the use of nanotube growth facilities. Financial support is obtained from the Japanese International Cooperative Research Project (ICORP) and the Dutch Fundamenteel Onderzoek der Materie (FOM).

**Figure 1.** Measurement scheme and basic sample characterization. **a,** Diagram showing the measurement circuit. Grey represents Ti/Al electrodes (10nm/60nm). Titanium ensures a good electrical contact to the CNT, while aluminium becomes superconducting below ~1.3K, well above the base temperature of our dilution refrigerator. The CNTs are probed in a four-terminal geometry (current bias, voltage measurement). An important element is the incorporation of three sets of filters for each measurement wire: a copper-powder filter (Cu-F) for high frequency noise, $\pi$-filters for intermediate frequencies and a two-stage RC filter to suppress voltage fluctuations at low frequencies. The dashed box region indicates the low temperature part of the circuit. The rest is at room temperature. **b,** Color-scale plot of the differential resistance, $dV/dI$, versus measured voltage, $V$, and gate voltage, $V_G$ at $T$=4.2K. The white arrow indicates the energy separation between discrete quantum levels in the CNT. **c,** Differential resistance versus measured source-drain voltage, $V$, at different temperatures (0.030, 0.47, 0.7, 0.88, 1.02, 1.18, 1.22 and 1.35 K, from bottom to top). The curves are offset for clarity (by 2k$\Omega$, for 0.47 and 0.7 K, and by 1k$\Omega$ for the rest). The features present in all curves below 1.3K are due to the induced superconducting proximity effect. The arrows indicate the superconducting gap at $V$=2$\Delta_g$/$e$~250$\mu$V. The dotted lines indicate multiple Andreev reflection (MAR) processes, which manifest as dips in $dV/dI$. **d,** $V(I)$-characteristics at base temperature showing the modulation of the critical current, $I_C$, with $V_G$ ($V_G$ = -2.59, -2.578, -2.57, -2.563, -2.555 and –2.541 V from black to orange). For currents larger than $I_C$ the system goes into a resistive state (abrupt jump from zero to finite $V$).

**Figure 2.** Quantum supercurrent transistor. **a,** Variation of the critical current, $I_C$, with gate voltage, $V_G$, extracted from **b** (see **c** and Fig. 3a for high resolution). $I_C$ is measured as the upper half-width of the black region around $I=0$. **b,** Color-scale representation (in logarithmic scale) of $dV/dI(I, V_G)$ at $T=30$mK (black is zero, i.e. supercurrent region, and $dV/dI$ increases from dark blue to white and red; the scale can be inferred from **d**). The differential resistance and critical current, $I_C$, exhibit a series of quasiperiodic modulations with gate voltage as the energy levels in the CNT QD are tuned ON and OFF resonance with respect to the Fermi energy in the superconducting leads. The sharp vertical features are caused by random charge switches and shift the diagram horizontally. The narrow tilted features present in the OFF regions (for example at $V_G \sim -2.87$V) occur reproducibly and are associated with Fano resonances[30] (see supplementary information). **c,** High-resolution $dV/dI(I, V_G)$ plot of the left-most resonance region in **b**. The modulation of $I_C$ (black central region) as well as multiple Andreev reflection (up to several orders, the first two are highlighted by the dashed blue lines) are clearly visible. **d,** Two representative $dV/dI$ ($I$) curves, taken from **c** at the vertical black and blue dashed lines, illustrating the different behaviour of the differential resistance in the ON (black curve/axis) and OFF (blue curve/axis) resonance case. **e,** Schematic diagram showing a strongly coupled QD in between two superconducting leads. The gate voltage tunes the position of the Lorentzian level from the ON (red curve) to the OFF (grey dashed curve) state.

**Figure 3.** Correlation between critical current and normal state conductance and modulation of the $I_C R_N$ product. In all panels, the black dots represent the experimental data points ($T=30$mK) and the red/blue curves are theoretical plots. **a,** Critical current, $I_C$, versus $V_G$ for the resonance shown in Fig. 2c. The theoretical lines are fits to $I_C = I_0 \left[1-(1-\Gamma_1\Gamma_2/((V_G-V_{GR})^2 + 0.25\Gamma^2))^{1/2}\right]$ (red curve) and $I_{CM} = I_{0M} \left[1-(1-\Gamma_1\Gamma_2/((V_G-V_{GR})^2 + 0.25\Gamma^2))^{1/2}\right]^{3/2}$ (blue), as explained in the main text. $V_{GR}$ is the value of gate voltage on resonance. All gate voltages and $\Gamma$'s are converted into energies by multiplying by the gate coupling factor, $\alpha=0.02$ meV/mV, obtained from measurements in the non-linear regime. **b**, Conductance, $G_N$, as a function of $V_G$ in the normal state ($B=40$mT) and the corresponding fit to $G_N = 4e^2/h$ $(\Gamma_1\Gamma_2/((V_G-V_{GR})^2 + 0.25\Gamma^2))$. **c,** $I_C$-$G_N$ correlation plot. The data show a non-trivial correlation, with a stronger decrease of $I_C$ than expected from the theoretical curve $I_C = I_0 [1-(1-¼G_N)^{1/2}]$ (red curve). The ¼ factor simply denotes that $G_N$ is measured in $e^2/h$ units. The difference can be almost entirely accounted for by the influence of the electromagnetic environment, resulting in a measured $I_{CM} = I_0 [1-(1-¼G_N)^{1/2}]^{3/2}$ (blue curve). An ideal SNS junction, with N a normal metal with continuous density of states, would exhibit a linear $I_C$-$G_N$ correlation curve (grey dashed curve). **d,** $I_C R_N$ product versus $V_G$, resulting from dividing the experimental data and theory curves from **a** and **b**. The grey dashed line indicates a constant $I_C R_N$ product such as in a SNS junction.

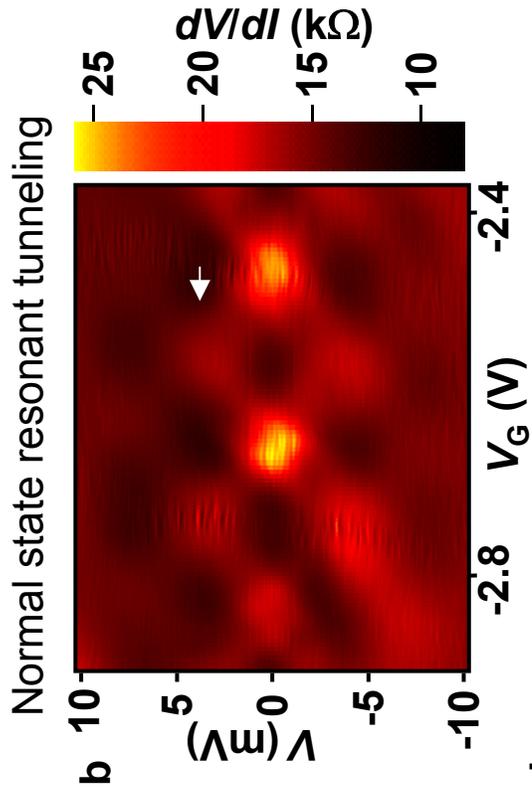

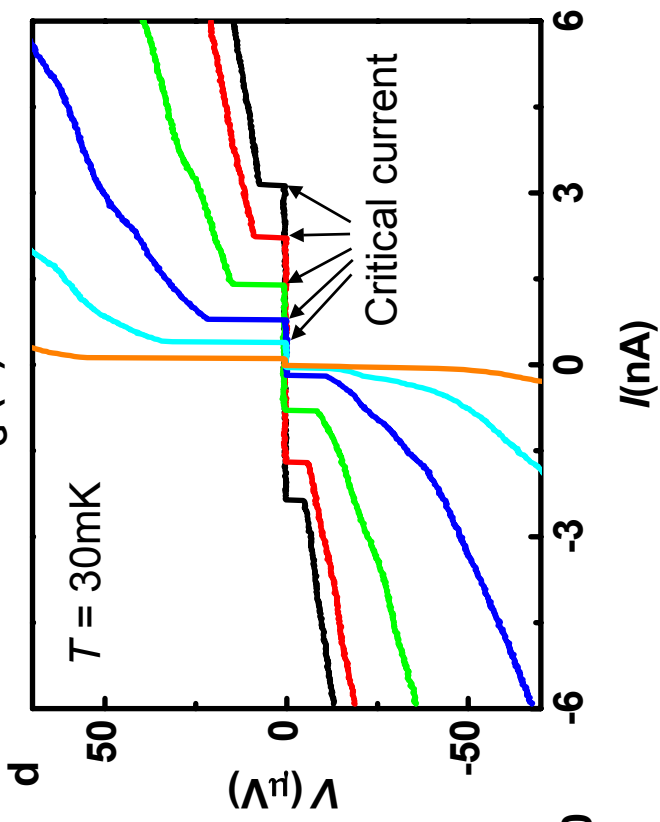

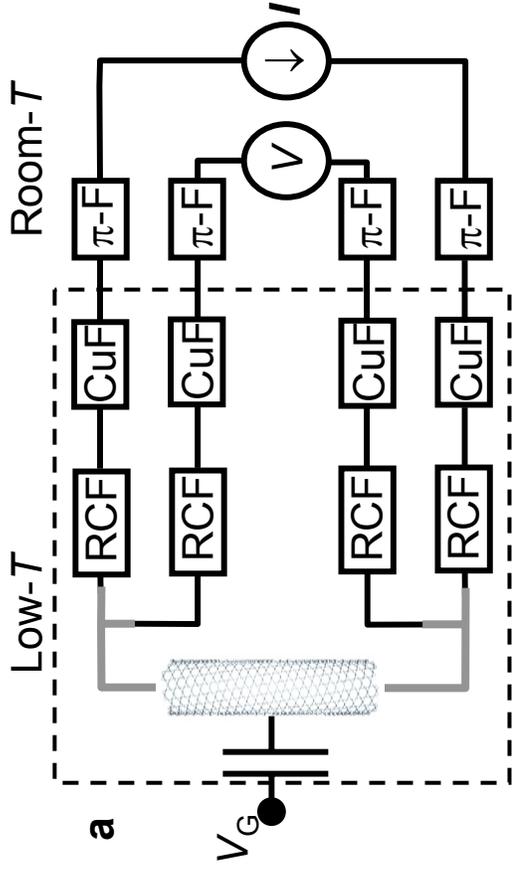

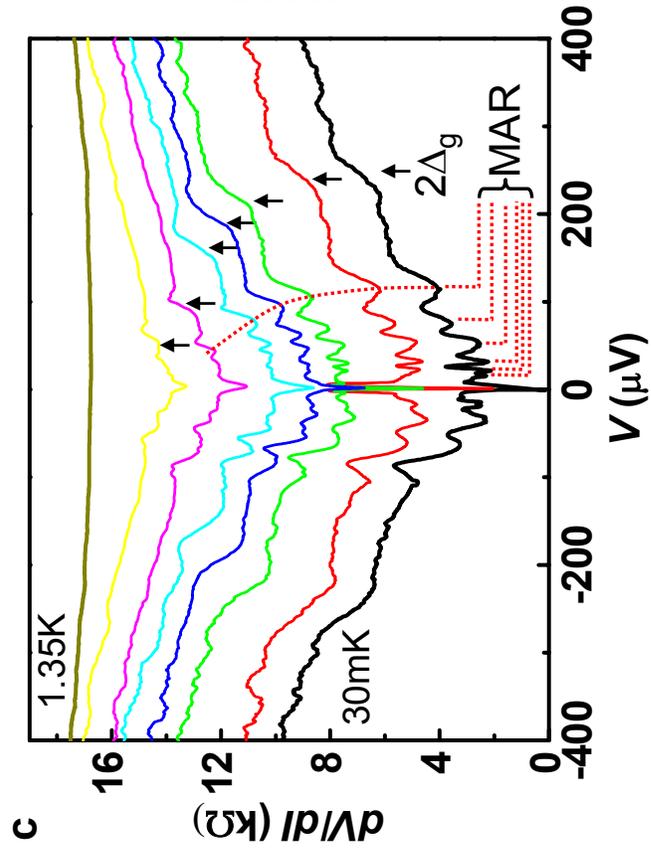

Figure 1 Jarillo-Herrero *et al.*

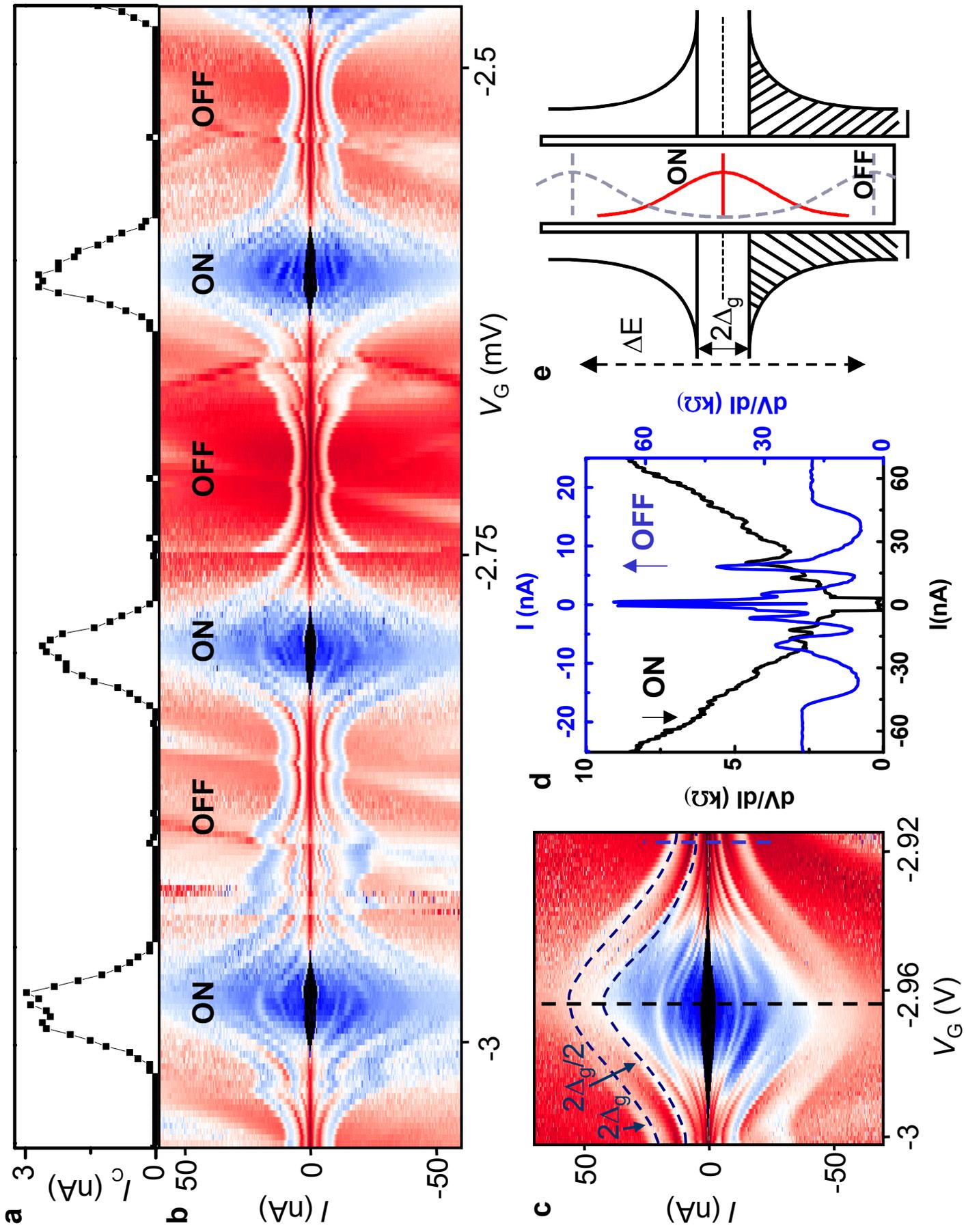

Figure 2 Jarillo-Herrero et al.

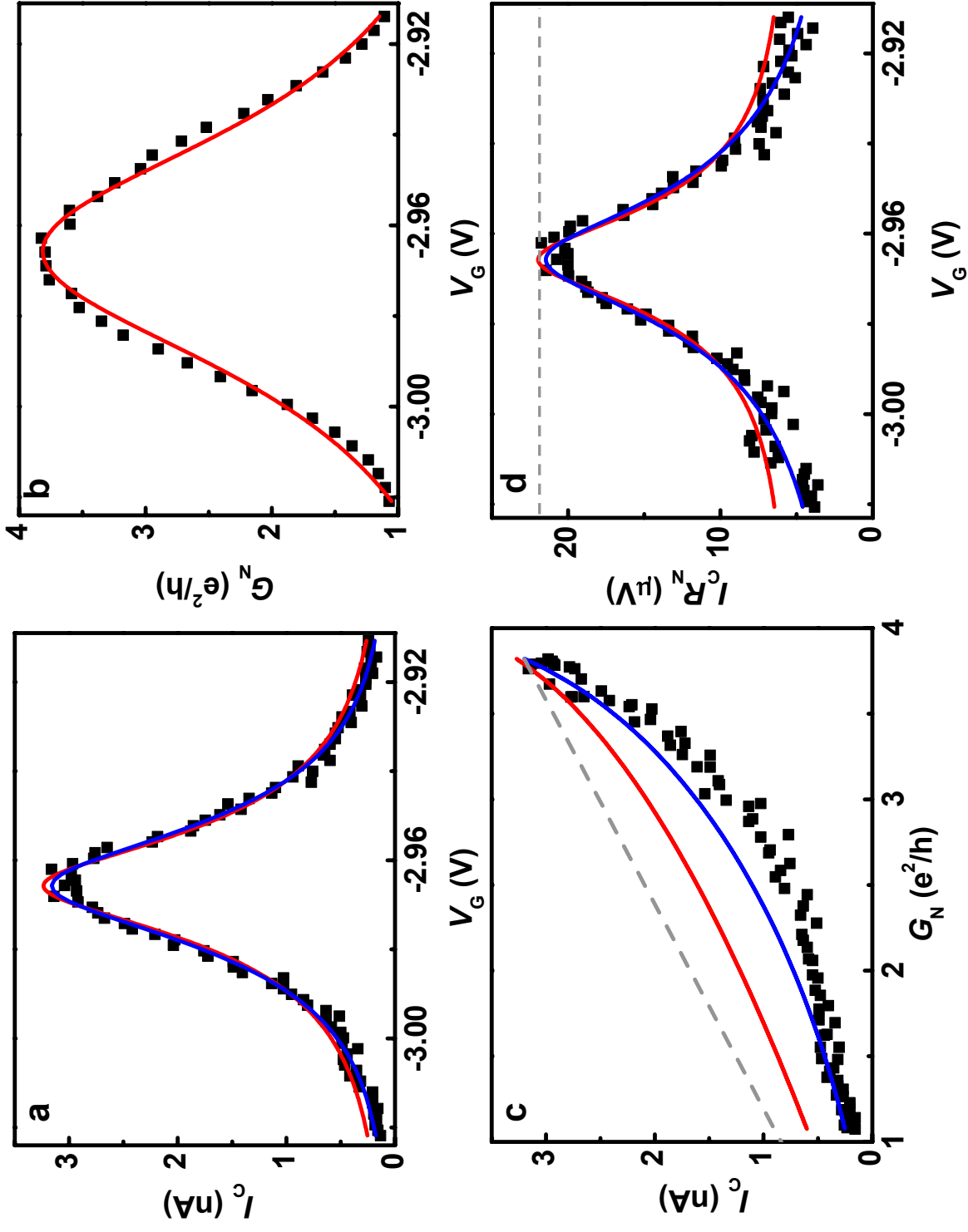

Figure 3 Jarillo-Herrero *et al.*